\documentclass[a4paper,11pt]{article}

\usepackage[dvipdfmx]{graphics}
\usepackage[dvipdfmx,svgnames]{xcolor}
\usepackage{latexsym,bm}
\usepackage{mathrsfs,amsmath,amssymb}
\usepackage{amsthm}

\usepackage{tikz}
\usetikzlibrary{decorations.pathmorphing}

\usepackage{bm}
\usepackage{here}

\def \inte #1 #2 {\int \limits_{#1}^{#2}}

\usepackage{amssymb}
\usepackage{amsmath}
\usepackage{color}

\usepackage{easywncy}

\usepackage{amsmath}

\usepackage{jheppub} 

\usepackage[T1]{fontenc} 

\def\localinput#1{{
  \renewcommand{\documentclass}[2][dummy]{}
  \renewcommand{\usepackage}[2][dummy]{}
  \renewenvironment{document}{}{}
  \def\jobname{#1}
  \input{#1}
}}

\def\sla#1{\ooalign{\hfil\hspace{-0.1ex}\raise.2ex\hbox{$\not \phantom{#1}$}\hfil\crcr  $#1$}}

\theoremstyle{plain}

\title{ The Fermi Function Beyond The Second Order Perturbation Theory}

\author[a]{Akihiro Matsuzaki, }
\emailAdd{akihiro@rikkyo.ac.jp }
\affiliation[a]{Tokai High School, \\Tutui 1-2-35, Higashi-ku, Nagoya-shi, Aichi 461-0003, Japan.}

\author[b]{Hidekazu Tanaka}%
\emailAdd{tanakah@rikkyo.ac.jp}
\affiliation[b]{Department of Physics, Rikkyo University,\\Nishi-Ikebukuro, Toshima-ku, Tokyo 171-8501, Japan.}%

\abstract{
    The Fermi function is historically derived from the Dirac equation or the Schr\"{o}dinger equation.
    However, we claim that the Fermi function should be derived from 
    quantum field theory. 
    Then, we obtain the following results:
    (1) We give the electromagnetic correction of the beta decay to the fourth order in $\alpha/v$, where $v$ is the electron velocity.
    (2) Our result is consistent with the ordinary non-relativistic Fermi function.
    (3)  We give the iterated integral representation of the Fermi function for arbitrary order in $\alpha/v$.
    (4) This representation is related to the multiple L values, and they enable us to calculate the amplitude in the algebraic approach.
}

\arxivnumber{1310.4082}
\begin{document}
\maketitle
\flushbottom

\section{Introduction}

    To evaluate the beta decay rates, for example $n\to p^+ +e^-+\bar \nu_e$, we introduce the Fermi function \cite{ZPhys88-161}, \cite{AmJP36-1150}, \cite{ProcRoySoc133-381}.
    It represents the effect of the electromagnetic potential caused by the proton. 
    The electron runs through this potential.  
    This function affects the beta spectrum, the decay width, and the lifetime of the parent particle \cite{NPA273-301}, \cite{NPA337-474}.
    The non-relativistic Fermi function has the form
\begin{align} \begin{split}\label{Eq-Fermi-Function}
F_{\mathrm{NR}}
=&\ \frac{2\pi\alpha/v}{1-e^{-2\pi\alpha/v}}
=1+\pi\left(\frac{\alpha}{v}\right)^1+\frac{\pi^2}{3}\left(\frac{\alpha}{v}\right)^2+0\left(\frac{\alpha}{v}\right)^3-\frac{\pi^4}{45}\left(\frac{\alpha}{v}\right)^4+\cdots,
\end{split} \end{align}
    where $\alpha$ is the fine structure constant and $v$ is the electron velocity relative to the daughter nucleon.  
    It is written as a function of $\alpha/v$.

    In our previous paper \cite{prc}, 
    we claimed that the Fermi function should be derived from the quantum field theory, and 
    we performed that the electromagnetic corrections derived from the quantum field theory are consistent with the non-relativistic Fermi function to the second order in $\alpha/v$. 
    Since the Fermi function originates from the electromagnetic interaction, 
    the effect of the Fermi function should be represented by 
    the diagrams involving the exchange of photons between electron and proton lines.

    From the theoretical point of view, we shall give eq. (\ref{Eq-Fermi-Function}) if we calculate the electromagnetic correction
    to all orders in $\alpha/v$. 
    It is not the simple exponential correction, unlike the sum of infrared corrections and the sum of disconnected diagrams. 
    It must be something more complicated. 
    We try to understand the properties of this summation. 

    The Fermi function, eq.(\ref{Eq-Fermi-Function}) is the generating function of the Bernoulli numbers. 
    The  Bernoulli numbers are represented by the zeta functions.
    On the other hand, the association of multiple zeta values with loop diagrams is pointed out \cite{Phys Lett B393-403}.
    We will show the relation between the Fermi function and the zeta functions.

    Already, some papers calculated the corrections to the Fermi function in quantum electrodynamics \cite{PR113-1652}, \cite{A. Sirlin}, \cite{PLB40-616}, 
    whereas we try to derive the Fermi function itself.

    In this paper, we give the non-relativistic Fermi function up to the fourth order in $\alpha/v$, and try to understand the general construction of the Fermi function.
    In Section \ref{nLoop}, we calculate the $n$th order diagram. 
    We explain the detail of calculation and we show the iterated integral representation, eq. (\ref{D}).
    In Section \ref{Third Order}, We show the result to the third order in $\alpha/v$.
    In Section \ref{MLV}, we point out that eq. (\ref{D}) is expressed by the multiple L values, and we calculate eq. (\ref{D}) in the algebraic approach, and show the result to the fourth order in $\alpha/v$. 
    In Section \ref{Conclusion}, we summarize our conclusions and give some discussion.

\section{$n$-Loop Ladder Diagrams}\label{nLoop}

    In the non-relativistic limit, the dominant contribution comes from the ladder diagram.\footnote{see Appendix \ref{TCD}.}
    We calculate the ladder diagram which is the $n$th order in $\alpha$.
    The amplitude is given by
\begin{align} \begin{split}\label{Eq8}
iM_n=&
\int\frac{d^4 k_1}{(2\pi)^4} \int\frac{d^4 k_2}{(2\pi)^4} \cdots \int\frac{d^4 k_n}{(2\pi)^4}
\bar{u}(q)
(-ie\gamma^{\mu_1})\frac{i\{(\sla q+\sla k_1)+m_e\}}{(q+k_1)^2-m_e^2+i\epsilon} \\
&\times
(-ie\gamma^{\mu_2})\frac{i\{(\sla q+\sla k_1+\sla k_2)+m_e\}}{(q+k_1+k_2)^2-m_e^2+i\epsilon} \cdots\\
&\times
(-ie\gamma^{\mu_n})\frac{i\{(\sla q+\sla k_1+\sla k_2+\cdots+\sla k_n)+m_e\}}{(q+k_1+k_2+\cdots+k_n)^2-m_e^2+i\epsilon}
P_L v(q') (-2\sqrt{2}iG_F)\\
&\times
\bar{u}(p)
(ie\gamma_{\mu_1})\frac{i\{(\sla p-\sla k_1)+m_p\}}{(p-k_1)^2-m_p^2+i\epsilon}
(ie\gamma_{\mu_2})\frac{i\{(\sla p-\sla k_1-\sla k_2)+m_p\}}{(p-k_1-k_2)^2-m_p^2+i\epsilon} \cdots\\
&\times
(ie\gamma_{\mu_n})\frac{i\{(\sla p-\sla k_1-\sla k_2-\cdots-\sla k_n)+m_p\}}{(p-k_1-k_2-\cdots-k_n)^2-m_p^2+i\epsilon}
P'_L u(p') \\
&\times
\frac{-i}{k_1^2-\mu^2+i\epsilon}
\frac{-i}{k_2^2-\mu^2+i\epsilon}\cdots
\frac{-i}{k_n^2-\mu^2+i\epsilon},\\
\end{split} \end{align}
    where $\mu$ is the infinitesimal photon mass, $\epsilon$ is an infinitesimal positive value, $e$ is the electric charge, $G_F$ is the Fermi constant; $u(p')$, $v(q')$, $\bar{u}(p)$, and $\bar{u}(q)$ represent the neutron, anti-neutrino, proton, and electron external lines, respectively, 
    and we define $\sla \ell=\ell^\mu \gamma_\mu$;  
    $P_L=(1- \gamma^5)/2$ and $P'_L=(1-C \gamma^5)/2$. $C$ represents the Gamow-Teller coupling constant relative to the Fermi constant.


    Since the dominant contribution to $iM_n$ comes from small $k_i$ in the non-relativistic limit, we eliminate $\sla {k_i}$.
    Then, we obtain 
\begin{align} \begin{split}\label{A}
&iM_n
\\
= &
(4ie^2 p\cdot q)^n iM_0 
\int\frac{d^4 k_1}{(2\pi)^4} \int\frac{d^4 k_2}{(2\pi)^4} \cdots \int\frac{d^4 k_n}{(2\pi)^4}
\frac{1}{k_1^2+2q\cdot k_1+i\epsilon}\cdots\\
&\times\frac{1}{(k_1+k_2+\cdots+k_n)^2+2q\cdot (k_1+k_2+\cdots+k_n)+i\epsilon}\\
&\times\frac{1}{k_1^2-2p\cdot k_1+i\epsilon}\cdots\\
&\times\frac{1}{(k_1+k_2+\cdots+k_n)^2-2p\cdot (k_1+k_2+\cdots+k_n)+i\epsilon}\\
&\times\frac{1}{k_1^2-\mu^2+i\epsilon}\frac{1}{k_2^2-\mu^2+i\epsilon}\cdots\frac{1}{k_n^2-\mu^2+i\epsilon},
\end{split} \end{align}
    where
\begin{align} \begin{split}
iM_0=-\frac{iG_F}{\sqrt{2}}\bar{u}(p)(1-C\gamma^5)u(p')\bar{u}(q)(1-\gamma^5)v(q')
\end{split} \end{align}
    is the tree level amplitude.

    Since $p_0 \gg q_0$, we approximate the denominators in eq. (\ref{A}) by 
\begin{align} \begin{split}
&\frac{1}{(k_1+k_2+\cdots+k_i)^2-2p\cdot (k_1+k_2+\cdots+k_i)+i\epsilon}\\
\simeq&
\frac{1}{-2p_0 (k_{10}+k_{20}+\cdots+k_{i0})+i\epsilon}.
\end{split} \end{align}
    We first integrate eq. (\ref{A}) over $k_{10}$. The $k_{10}$ integral can be performed as a contour integral in the complex plane. 
    The poles are at $k_{10}\simeq 0$, $k_{10}\simeq q_0$, and $k_{10}\simeq \sqrt{\bm{k}_1^2+\mu^2}$.
    Since the dominant contribution comes from small $k_i$ in the non-relativistic limit, we close the contour upward, picking up only the pole at $k_{10}\simeq 0$. 
    Thus, the amplitude is 
\begin{align} \begin{split}
iM_n \hspace{2cm}&\\
=\left(\frac{\alpha q_0}{\pi^2}\right)^n iM_0 &\int d^3 \bm{k}_1 \int d^3 \bm{k}_2 \cdots \int d^3 \bm{k}_n 
\frac{1}{\bm{k}_1^2+\mu^2} \frac{1}{\bm{k}_2^2+\mu^2}\cdots\frac{1}{\bm{k}_n^2+\mu^2}\\
&\times\frac{1}{\bm{k}_1^2+2\bm{q}\cdot \bm{k}_1-i\epsilon} 
\frac{1}{(\bm{k}_1+\bm{k}_2)^2+2\bm{q}\cdot(\bm{k}_1+ \bm{k}_2)-i\epsilon}\cdots\\
&\times\frac{1}{(\bm{k}_1+\bm{k}_2+\cdots+\bm{k}_n)^2+2\bm{q}\cdot (\bm{k}_1+\bm{k}_2+\cdots+\bm{k}_n)-i\epsilon}\\
=\left(\frac{\alpha q_0}{\pi^2}\right)^n iM_0 &
\prod_{k=1}^n \int d^3 \bm{k}_k  \frac{1}{\bm{k}_k^2+\mu^2} \frac{1}{(\sum_{i=1}^k \bm{k}_i)^2+2\bm{q}\cdot (\sum_{i=1}^k \bm{k}_i)-i\epsilon}.
\end{split} \end{align}
    For simplicity, we define $\bm{x}_i=\bm{k}_i/|\bm{q}|$, $ \bar{\mu}=\mu/|\bm{q}|$, $\hat{\bm{q}}=\bm{q}/|\bm{q}|$, $x_k=|\bm{x}_k|$. 
    Then, we obtain 
\begin{align} \begin{split}
iM_n 
=&\left(\frac{\alpha q_0}{\pi^2 |\bm{q}|}\right)^n iM_0 
\prod_{k=1}^n \int d^3 \bm{x}_k  \frac{1}{x_k^2+\bar{\mu}^2} \frac{1}{(\sum_{ i=1}^k \bm{x}_i +\hat{\bm{q}})^2-1-i\epsilon}.
\end{split} \end{align}
    Here, we define 
\begin{align} \begin{split}
t_k=\left|\sum_{i=1}^k \bm{x}_i+\hat{\bm{q}}\right|
=\sqrt{x_k^2+\left|\sum_{i=1}^{k-1} \bm{x}_i+\hat{\bm{q}}\right|^2+2x_k\left|\sum_{i=1}^{k-1} \bm{x}_i+\hat{\bm{q}}\right|\cos\theta_k},
\end{split} \end{align}
    where $\theta_k$ is the angle between $\bm{x}_k$ and $(\sum_{i=1}^{k-1} \bm{x}_i+\hat{\bm{q}})$.
    We set $t_0=|\hat{\bm{q}}|=1$.
    Introducing the spherical coordinates, we have
\begin{align} \begin{split}
d\cos\theta_k=\frac{1}{x_k}\frac{t_k}{t_{k-1}}dt_k
\end{split} \end{align}
and
\begin{align} \begin{split}
iM_n 
=\left(\frac{2\alpha }{\pi v}\right)^n iM_0 &
\prod_{k=1}^n  \int^{\infty}_{0}dx_k \inte{|x_k-t_{k-1}|} {x_k+t_{k-1}} dt_k \frac{x_k}{x_k^2+\bar{\mu}^2} \frac{1}{t_{k-1}}\frac{t_k}{t_k^2-1-i\epsilon} \\
=\left(\frac{2\alpha }{\pi v}\right)^n iM_0 &
 \left(\prod_{k=1}^{n-1} \int^{\infty}_{0}dx_k \inte{|x_k-t_{k-1}|} {x_k+t_{k-1}} dt_k \frac{x_k}{x_k^2+\bar{\mu}^2} \frac{1}{t_k^2-1-i\epsilon}\right)\\
&\ \ \qquad \times\int^{\infty}_{0}dx_n \inte{|x_n-t_{n-1}|} {x_n+t_{n-1}} dt_n \frac{x_n}{x_n^2+\bar{\mu}^2} \frac{t_n}{t_n^2-1-i\epsilon}.
\end{split} \end{align}

    Changing the order of integration, it becomes
\begin{align} \begin{split}
iM_n 
=&\left(\frac{2\alpha }{\pi v}\right)^n iM_0 
 \left(\prod_{k=1}^{n-1} \int^{\infty}_{0}dt_k \inte{|t_k-t_{k-1}|} {t_k+t_{k-1}} dx_k \frac{x_k}{x_k^2+\bar{\mu}^2} \frac{1}{t_k^2-1-i\epsilon}\right)\\
&\times\int^{\infty}_{0}dt_n \inte{|t_n-t_{n-1}|} {t_n+t_{n-1}} dx_n \frac{x_n}{x_n^2+\bar{\mu}^2} \frac{t_n}{t_n^2-1-i\epsilon}\\
=&\left(\frac{\alpha }{\pi v}\right)^n iM_0 
 \Biggl(\prod_{k=1}^{n} \int^{\infty}_{0}dt_k \frac{1}{t_k^2-1-i\epsilon}\\
&\hspace{0.4cm}\times \left[\log\{(t_k+t_{k-1})^2+\bar{\mu}^2\}-\log\{(t_k-t_{k-1})^2+\bar{\mu}^2\} \right] \Biggr)\times t_n.
\end{split} \end{align}

    Since the integrand is symmetric with respect to $t_k\to-t_k$, 
    we have 
\begin{align} \begin{split}
&\int^{\infty}_{0}dt_k \frac{1}{t_k^2-1-i\epsilon}\left[\log\{(t_k+t_{k-1})^2+\bar{\mu}^2\}-\log\{(t_k-t_{k-1})^2+\bar{\mu}^2\} \right]\\
=&\int^{\infty}_{0}dt_k \frac{1}{t_k^2-1-i\epsilon}\left[\log\frac{t_k+t_{k-1}+i\bar{\mu}}{t_k-t_{k-1}+i\bar{\mu}}+\log\frac{t_k+t_{k-1}-i\bar{\mu}}{t_k-t_{k-1}-i\bar{\mu}} \right]\\
=&\int^{\infty}_{-\infty}dt_k \frac{1}{t_k^2-1-i\epsilon}\log\frac{t_k+t_{k-1}+i\bar{\mu}}{t_k-t_{k-1}+i\bar{\mu}}.
\end{split} \end{align}
    The amplitude becomes
\begin{align} \begin{split}
iM_n=
\left(\frac{\alpha }{\pi v}\right)^n iM_0 
\left(\prod_{k=1}^{n} \int^{\infty}_{-\infty}dt_k \frac{1}{t_k^2-1-i\epsilon}\log\frac{t_k+t_{k-1}+i\bar{\mu}}{t_k-t_{k-1}+i\bar{\mu}}  \right)\times t_n.
\end{split} \end{align}

    For $k \ge1$, we define
\begin{align} \begin{split}\label{B}
F_k=&
\int^{\infty}_{-\infty} dt_k     \frac{1}{t_k^2    -1-i\epsilon}\log\frac{t_k+1+i\bar{\mu}}{t_k-t_{k-1}+i\bar{\mu}}
\int^{\infty}_{-\infty} dt_{k+1} \frac{1}{t_{k+1}^2-1-i\epsilon}\log\frac{t_{k+1}+t_k+i\bar{\mu}}{t_{k+1}-t_k+i\bar{\mu}}.
\end{split} \end{align}
    The $t_k$ integral can be performed as a contour integral in the complex plane. 
    We close the contour upward for $\log(t_{k+1}+t_k+i\bar{\mu})$, and
    we close the contour downward for $\log(t_{k+1}-t_k+i\bar{\mu})$. 
    We obtain     
\begin{align} \begin{split}
F_k=&
\int^{\infty}_{-\infty}dt_{k+1} \frac{1}{t_{k+1}^2-1-i\epsilon} \\
&\times\left\{
\frac{2\pi i}{2}\log\frac{2+i\bar{\mu}}{1-t_{k-1}+i\bar{\mu}} \log(t_{k+1}+1+i\bar{\mu}) \right.\\
&\ \ -\frac{2\pi i}{2}\log\frac{i\bar{\mu}}{-1-t_{k-1}+i\bar{\mu}} \log(t_{k+1}+1+i\bar{\mu})\\
&\ \ \left. +2\pi i \inte {-1-i\bar{\mu}} {t_{k-1}-i\bar{\mu}} dt_k \frac{1}{t_k^2-1}\log(t_{k+1}-t_k+i\bar{\mu})\right\}\\
=&
(-2\pi i)
 \inte {-1-i\bar{\mu}} {t_{k-1}-i\bar{\mu}} dt_k \frac{1}{t_k^2-1}
\int^{\infty}_{-\infty} dt_{k+1} \frac{1}{t_{k+1}^2-1-i\epsilon} 
\log\frac{t_{k+1}+1+i\bar{\mu}}{t_{k+1}-t_k+i\bar{\mu}}.
\end{split} \end{align}
    The last line is expressed by the integral. 
    Substituting this formula from $k=1$ to $k=n-1$ in order, we have
\begin{align} \begin{split}
iM_n
= &
\left(\frac{\alpha }{\pi v}\right)^n iM_0 (-2\pi i)^{n-1}
\left(\prod_{k=1}^{n-1} \inte {-1-i\bar{\mu}} {t_{k-1}-i\bar{\mu}} dt_k \frac{1}{t_k^2-1}\right) \\
&\times \int^{\infty}_{-\infty}dt_n \frac{t_n}{t_n^2-1-i\epsilon}\log\frac{t_n+1+i\bar{\mu}}{t_n-t_{n-1}+i\bar{\mu}}. 
\end{split} \end{align}
    Since the integrand vanishes faster than $t_n^{-1}$ for $|t_n|\to \infty$, 
    we close the contour upward, and obtain
\begin{align} \begin{split}
&\int^{\infty}_{-\infty}dt_n \frac{t_n}{t_n^2-1-i\epsilon}\log\frac{t_n+1+i\bar{\mu}}{t_n-t_{n-1}+i\bar{\mu}} \\
=&\frac{2\pi i}{2}\log\frac{2+i\bar{\mu}}{1-t_{n-1}+i\bar{\mu}}
=\frac{-2\pi i}{2} \inte {-1-i\bar{\mu}} {t_{n-1}-i\bar{\mu}} dt_n \frac{1}{t_n-1}.
\end{split} \end{align}
    Then, the amplitude is written as
\begin{align} \begin{split}
iM_n
= &
\frac{1}{2}\left(\frac{-2i\alpha }{ v}\right)^n iM_0 
\left(\prod_{k=1}^{n-1} \inte {-1-i\bar{\mu}} {t_{k-1}-i\bar{\mu}} dt_k \frac{1}{t_k^2-1}\right) 
\inte {-1-i\bar{\mu}} {t_{n-1}-i\bar{\mu}} dt_n \frac{1}{t_n-1} \\
= &
\left(\frac{i\alpha }{ v}\right)^n iM_0 
\left\{\prod_{k=1}^{n-1} \inte {(k-1)i\frac{\bar{\mu}}{2}} {t_{k-1}'} dt_k' \left(\frac{1}{1-t_k'+ik\frac{\bar{\mu}}{2}}+\frac{1}{t_k'-ki\frac{\bar{\mu}}{2}}\right)\right\} \\
&\hspace{2cm}\times\inte {(n-1)i\frac{\bar{\mu}}{2}} {t_{n-1}'} dt_n' \frac{1}{1-t_n'+ni\frac{\bar{\mu}}{2}}, 
\end{split} \end{align}
    where $t_k'=(1+t_k+ik\bar{\mu})/2$ and $k=1,2,\ldots,n$.
    In the limit $\bar{\mu}\to0$, this can be written as
\begin{align} \begin{split}\label{D}
iM_n
= &
\lim_{\bar \mu\to0}
\left(\frac{i\alpha }{ v}\right)^n iM_0 
\left\{\prod_{k=1}^{n-1} \int^{t_{k-1}}_{0}dt_k \left(\frac{1}{1+ik\frac{\bar{\mu}}{2}-t_k}+\frac{1}{t_k}\right)\right\} \\
&\hspace{2cm}\times\int^{t_{n-1}}_{0}dt_n \frac{1}{1+ni\frac{\bar{\mu}}{2}-t_n}, 
\end{split} \end{align}
    where $t_0=1$.
    This is the iterated integral representation of the amplitude of the electromagnetic correction for the $n$th order in $\alpha$.

\section{The Third-Order Amplitude}\label{Third Order}

    For $n=3$ case, we obtain the following result:
    The amplitude is 
\begin{align} \begin{split}\label{C}
iM_3
= 
\left(\frac{\alpha}{v}\right)^3 iM_0 
\left[
-\frac{\pi^3}{48}-\frac{\pi}{4}\log^2 \frac{\bar{\mu}}{2}
+\frac{i}{24}
\left\{
-\pi^2\log \frac{\bar{\mu}}{2}+4\log^3 \frac{\bar{\mu}}{2}-8\zeta(3)
\right\}
\right],
\end{split} \end{align}
    where we use the formulae, 
\begin{align} \begin{split}
L_2(x)=&\int^{x}_{0} \frac{-\log(1-t)}{t}\ dt,\\
L_3(x)=&\int^{x}_{0} \frac{L_2(t)}{t}\ dt,\\
L_2\left(\frac{1}{9}\right)=&-\frac{\pi^2}{3}+\log^2 3+6L_2\left(\frac{1}{3}\right),\\
L_2\left(\frac{1}{3}\right)=&-\frac{1}{2}\log^2 \frac{2}{3}-L_2\left(-\frac{1}{2}\right),\\
L_3(3)=&-\frac{i\pi}{2}\log^2 3+\frac{\pi^2}{6}\log9-\frac{1}{6}\log^3 3+L_3\left(\frac{1}{3}\right),\\
L_3\left(\frac{1}{9}\right)=&-\frac{2}{3}\log^3 3+\frac{\pi^2}{3}\log9+12L_3\left(\frac{1}{3}\right)-\frac{26}{3}\zeta(3),\\
\end{split} \end{align}
    where $\zeta(3)$ is a Riemann zeta function with index 3. 
    For $n=1$ and $2$, it is easy to calculate the amplitudes $iM_1$ and $iM_2$.
    Then, the amplitude up to the third order in $\alpha$ is  
\begin{align} \begin{split}\label{E3}
\sum_{k=0}^{3} (iM_k)=&iM_0\left\{1+\frac{\alpha}{v}\left(\frac{\pi}{2}-i\log\frac{\bar{\mu}}{2}\right)
-\frac{\alpha^2  }{v^2}
\left(-\frac{1}{24}\pi^2+\frac{1}{2}\log^2\frac{\bar{\mu}}{2}+\frac{i\pi}{2}\log\frac{\bar{\mu}}{2}\right)\right\}
+iM_3.\\
\end{split} \end{align}
    The first and second order terms in eq. (\ref{E3}) were already derived in Ref. \cite{prc}.

    The absolute square of eq. (\ref{E3}) is
\begin{align} \begin{split}\label{E4}
\left|\sum_{k=0}^{3} (iM_k)\right|^2=&|M_0|^2 \left\{ 1+\pi\left(\frac{\alpha}{v}\right)^1+\frac{\pi^2}{3}\left(\frac{\alpha}{v}\right)^2+0\left(\frac{\alpha}{v}\right)^3 \right\}+\mathcal{O}\bigl((\alpha/v)^4\bigr).
\end{split} \end{align}
    In the above expression, the infrared divergences cancel between the ladder diagrams up to the third order in $\alpha/v$.
    Eq. (\ref{E4}) is consistent with the non-relativistic formula of the Fermi function (\ref{Eq-Fermi-Function}). 
    The third order terms in eq. (\ref{E4}) are canceled as the expansion of the Fermi function are.
    It is non-trivial.
    However, it is difficult to calculate the amplitude for $n\ge4$. 

\section{Multiple L Values }\label{MLV}

    In this section, we try to calculate eq.(\ref{D}) algebraically.
    This approach is more scalable than the method shown in the previous section.

    In Ref. \cite{Arakawa1}, two types of multiple L values, $L_\mathcyr{x}$ and $L_*$ are defined as
\begin{align} \begin{split}\label{Def1}
L_\mathcyr{x}(k_1,&\ldots,k_n;a_1,\ldots,a_n)\\
&=
\sum_{m_1> m_2>\cdots> m_n\ge 1}
\frac{\zeta^{a_1 (m_1-m_2)}\zeta^{a_2 (m_2-m_3)}\cdots \zeta^{a_{n-1} (m_{n-1}-m_n)} \zeta^{a_n m_n}}{m_1^{k_1} m_2^{k_2} \cdots m_n^{k_n}},
\\
L_*(k_1,&\ldots,k_n;a_1,\ldots,a_n)\\
&=
\sum_{m_1> m_2>\cdots> m_n\ge 1}
\frac{\zeta^{a_1 m_1}\zeta^{a_2 m_2}\cdots \zeta^{a_n m_n}}{m_1^{k_1} m_2^{k_2} \cdots m_n^{k_n}},
\end{split} \end{align}
    where $k_1,\ldots,k_n$ are positive integers, $a_1,\ldots,a_n $ are integers, and $\zeta=\exp[2\pi i/m]$, where $m$ is a natural number.
    The multiple L values are the generalizations of the multiple zeta values.

    The series $L_\mathcyr{x}(k_1,\ldots,k_n;a_1,\ldots,a_n)$ has an integral expression as follows.
\begin{align} \begin{split}\label{MLSV}
&L_\mathcyr{x}(k_1,\ldots,k_n;a_1,\ldots,a_n)\\
= &
\underbrace{\int_0^t   \frac{dt}{t} \cdots \int_0^t \frac{dt}{t} }_{k_1-1} 
 \int_0^t \frac{\zeta^{a_1}}{1-\zeta^{a_1}t}
 \cdots 
\underbrace{\int_0^t   \frac{dt}{t} \cdots \int_0^t \frac{dt}{t} }_{k_n-1} 
\int_0^t dt \frac{\zeta^{a_n}}{1-\zeta^{a_n}t}. 
\end{split} \end{align}

    For $k\bar{\mu}\ll 1$, we approximate $(1+ik\frac{\bar{\mu}}{2}) \simeq \exp[ik\frac{\bar{\mu}}{2}]=(\exp[i\frac{\bar{\mu}}{2}])^{k}$.
     Let $m$ be a natural number such that $\bar{\mu}=4\pi/m$.
     Then, $\zeta
     = \exp[i\frac{\bar{\mu}}{2}]$, and we obtain
\begin{align} \begin{split}
\frac{iM_n}{ iM_0 }\left(\frac{i\alpha }{ v}\right)^{-n}
= &
\left\{\prod_{k=1}^{n-1} \int^{t_{k-1}}_{0}dt_k \left(\frac{\zeta^{-k}}{1-\zeta^{-k} t_k}+\frac{1}{t_k}\right)\right\} 
\int^{t_{n-1}}_{0}dt_n \frac{\zeta^{-n}}{1-\zeta^{-n}t_n}. 
\end{split} \end{align}
    For $n=1$, we have
\begin{align} \begin{split}
\frac{iM_1}{ iM_0}\left(\frac{i\alpha }{ v}\right)^{ -1}
= &
\int^{t_{0}}_{0}dt_1 \frac{\zeta^{-1}}{1-\zeta^{-1}t_1}
=
L_\mathcyr{x}(1;-1) 
=
L_*(1;-1). 
\end{split} \end{align}
    For $n=2$, we have
\begin{align} \begin{split}
&\frac{iM_2}{ iM_0}\left(\frac{i\alpha }{ v}\right)^{-2}\\
= &
 \int^{t_{0}}_{0}dt_1 \left(\frac{\zeta^{-1}}{1-\zeta^{-1} t_1}+\frac{1}{t_1}\right)
\int^{t_{1}}_{0}dt_2 \frac{\zeta^{-2}}{1-\zeta^{-2}t_2}\\
=&
L_\mathcyr{x}(1,1;-1,-2)+L_\mathcyr{x}(2;-2) \\
=&
L_*(1,1;-1,-1)+L_*(2;-2)
=\frac{1}{2}
\left\{L_*(1;-1)^2+L_*(2;-2)\right\}
. 
\end{split} \end{align}
    In the last step, we use the relations;    
\begin{align} \begin{split}\label{L11}
&L_*(1,1;-1,-1)\\
=&
\sum_{m_1> m_2}
\frac{\zeta^{- m_1}\zeta^{- m_2}}{m_1 m_2}
=
\frac{1}{2}\left\{
\sum_{m_1> m_2}
\frac{\zeta^{- m_1}\zeta^{- m_2}}{m_1 m_2}
+\sum_{m_1< m_2}
\frac{\zeta^{- m_1}\zeta^{- m_2}}{m_1 m_2}
+\sum_{m_1}
\frac{\zeta^{-2 m_1}}{m_1^2 }
\right\}
-\frac{1}{2}\sum_{m_1}
\frac{\zeta^{-2 m_1}}{m_1^2 }
\\
=&\frac{1}{2}
\sum_{m_1}\sum_{m_2}
\frac{\zeta^{- m_1}}{m_1 }
\frac{\zeta^{- m_2}}{ m_2}
-\frac{1}{2}\sum_{m_1}
\frac{\zeta^{-2 m_1}}{m_1^2 }
=\frac{1}{2}\left\{
L_*(1;-1)^2-L_*(2;-2)
\right\}.\\
\end{split} \end{align}
    For $n=3$, we have
\begin{align} \begin{split}
&\frac{iM_3}{ iM_0}\left(\frac{i\alpha }{ v}\right)^{-3}\\
= &
\int^{t_{0}}_{0}dt_1 \left(\frac{\zeta^{-1}}{1-\zeta^{-1} t_1}+\frac{1}{t_1}\right) 
\int^{t_{1}}_{0}dt_2 \left(\frac{\zeta^{-2}}{1-\zeta^{-2} t_2}+\frac{1}{t_2}\right)
\int^{t_{2}}_{0}dt_3 \frac{\zeta^{-3}}{1-\zeta^{-3}t_3}\\
=&
L_\mathcyr{x}(1,1,1;-1,-2,-3)+L_\mathcyr{x}(2,1;-2,-3)+L_\mathcyr{x}(1,2;-1,-3)+L_\mathcyr{x}(3;-3) 
\\
=&
L_*(1,1,1;-1,-1,-1)+L_*(2,1;-2,-1)+L_*(1,2;-1,-2)+L_*(3;-3) 
\\
=&\frac{1}{6}
\left\{L_*(1;-1)^3+3L_*(1;-1)L_*(2;-2)+2L_*(3;-3)\right\}.
\end{split} \end{align}
    In the last line, we use the following relations;
\begin{align} \begin{split}
&L_*(1,2;-1,-2)+L_*(2,1;-2,-1)+L_*(3;-3)
=L_*(1;-1)L_*(2;-2),\\
&6L_*(1,1,1;-1,-1,-1)+3L_*(1,2;-1,-2)+3L_*(2,1;-2,-1)+L_*(3;-3)
=L_*(1;-1)^3.
\end{split} \end{align}
    The derivation is similar to eq. (\ref{L11}).
    For $n=4$,
\begin{align} \begin{split}
\frac{iM_4}{ iM_0}\left(\frac{i\alpha }{ v}\right)^{-4}
=&\frac{1}{24} 
\bigl\{L_*(1;-1)^4+6L_*(1;-1)^2 L_*(2;-2)\\
&+8L_*(1;-1)L_*(3;-3)+3L_*(2;-2)^2+6L_*(4;-4)\bigr\}.
\end{split} \end{align}

    Then, the absolute square of the amplitude to fourth order in $\alpha/v$ is  
\begin{align} \begin{split}\label{Eq35}
&\left |\sum_{k=0}^4 (iM_k)\right|^2\\
=&|iM_0|^2 
\Bigl[
1+\pi\left(\frac{\alpha }{ v}\right)^1 
+ \frac{\pi^2}{3}\left(\frac{\alpha }{ v}\right)^2
+0\left(\frac{\alpha }{ v}\right)^3
-\frac{\pi^4}{45}\left(\frac{\alpha }{ v}\right)^4
\Bigr]   
+\mathcal{O}\bigl((\alpha/v)^5\bigr).
\end{split} \end{align}
    For $n\ge 2$, $L_*(n;-n)$ and $\bar L_*(n;-n)$ are finite in the limit $\bar\mu\to0$ as 
\begin{align} \begin{split}
\lim_{\bar \mu \to0}L_*(n;-n)&=\lim_{\bar \mu \to0} \bar L_*(n;-n)=\zeta(n).
\end{split} \end{align}
%
%
%
%
    The infrared divergences arise from only $L_*(1;-1)$ and $\bar L_*(1;-1)$.
    Each of them has a divergence in the same limit. 
    However, only their difference appears in the absolute square. 
    Their difference,   
\begin{align} \begin{split}
\lim_{\bar \mu \to0}\{L_*(1;-1)-\bar L_*(1;-1)\}=&\lim_{\bar\mu\to0}\left(\int^{1}_{0}dt \frac{\zeta^{-1}}{1-\zeta^{-1}t}-\int^{1}_{0}dt \frac{\zeta}{1-\zeta t}\right)
=-i\pi 
\end{split} \end{align}
    has a finite value.

    Eq. (\ref{Eq35}) is consistent with eq. (\ref{Eq-Fermi-Function}).

\section{ The Concluding Remarks }\label{Conclusion}

    We showed the non-relativistic corrections of beta-decay amplitude are represented by the iterated integral, eq. (\ref{D}).
    In Section \ref{Third Order}, we performed the electromagnetic corrections to the third order in $\alpha /v$.
    It is consistent with the non-relativistic Fermi function.
    However, it is practically difficult to calculate the amplitude for $n\ge 3$.
    We introduced the two types of multiple L values, $L_\mathcyr{x}$ and $L_*$.
    eq. (\ref{D}) is written in terms of the L values.
    They gave us an algebraic approach to calculate the amplitude, (\ref{D}).
    The cancellation of the infrared divergence is explicit in this approach.
    We showed the electromagnetic corrections to the fourth order in $\alpha /v$, and it is also consistent with the non-relativistic Fermi function.
    We confirmed that the Fermi function is derived from the quantum field theory up to the fourth order in $\alpha$.

    We derived the Fermi function in the non-relativistic approximation.
    However, the quantum electrodynamics itself is relativistic theory.
    We can consider the relation between non-relativistic limit and the relativistic limit.
    In the relativistic limit, $\alpha/v$ and $\alpha$ have the same value, 
    and we have no reason to distinguish the Fermi function and other corrections. 
    We can say that the Fermi function is the electromagnetic correction which is dominant in the non-relativistic limit.


    Our result suggests that the absolute square of eq. (\ref{D}) leads to the Fermi function (\ref{Eq-Fermi-Function}).
    It is practically difficult for higher order to calculate the amplitudes.
    We should develop the systematic method for calculating the higher order amplitudes.
    This paper must be a significant milestone to reach them.
    We will present it, soon.



\appendix

\section{ The Dominant Contribution}\label{TCD}

\begin{figure}[tb]
  \begin{center}
    \includegraphics[keepaspectratio=true,height=25mm]{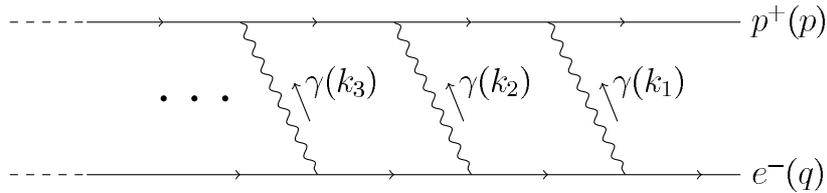}
  \end{center}
  \caption{One of the time ordered ladder diagrams. Time flies from left to right.}
  \label{Ladder}
\end{figure}

    We use the old-fashioned perturbation theory since it is convenient to see where the infrared divergences arise from.

    We search for the diagrams which contribute to the electromagnetic correction for the $n$th order in $\alpha/v$.
    We prove that only the ladder diagrams contribute to the electromagnetic correction in the old-fashioned perturbation theory \cite{Weinberg}.
 
    First, we consider the ladder diagrams.
    The situation "A proton radiates a photon and it is absorbed by an electron, then the proton radiates another photon and the electron absorbs it, and so on"
    is shown in figure \ref{Ladder}.
    This diagram corresponds to the amplitude 
\begin{align} \begin{split}\label{E1}
iM_n \propto&
\int d^3\bm{k}_1
\frac{m_p q^0}{\omega_\gamma(\bm{k}_1) \omega_p(-\bm{k}_1) \omega_e(\bm{q}+\bm{k}_1)}
\frac{1}{\omega_e(\bm{q})-\omega_e(\bm{q}+\bm{k}_1)-\omega_\gamma(\bm{k}_1) }\\
&\times
\frac{1}{\omega_p(\bm{0})+ \omega_e(\bm{q})-\omega_p(-\bm{k}_1)- \omega_e(\bm{q}+\bm{k}_1)}\\
&\times
\int d^3\bm{k}_2
\frac{m_p q^0}{\omega_\gamma(\bm{k}_2) \omega_p(-\bm{k}_1-\bm{k}_2) \omega_e(\bm{q}+\bm{k}_1+\bm{k}_2)}\\
&\times
\frac{1}{\omega_p(\bm{0})+\omega_e(\bm{q})-\omega_p(-\bm{k}_1)-\omega_e(\bm{q}+\bm{k}_1+\bm{k}_2)-\omega_\gamma(\bm{k}_2) }\\
&\times
\frac{1}{\omega_p(\bm{0})+ \omega_e(\bm{q})-\omega_p(-\bm{k}_1-\bm{k}_2)- \omega_e(\bm{q}+\bm{k}_1+\bm{k}_2)}\\
&\times \cdots
\end{split} \end{align}
    for the rest frame of the final state proton,
    where the external momenta of proton and electron are $\bm{p}$ and $\bm{q}$, respectively; the momenta of photons are $\bm{k_i}$; $m_p$ is the proton mass; $m_e$ is the electron mass; $\omega_p(\bm{\ell})= \sqrt{\bm{\ell}^2+m_p^2}$, $\omega_e(\bm{\ell})= \sqrt{\bm{\ell}^2+m_e^2}$, and $\omega_\gamma(\bm{\ell})= |\bm{\ell}|$ represent the proton energy, electron energy, and the photon energy, respectively; $q^0=\omega_e(\bm{q})$ and $p^0=\omega_p(0)$ are the final-state electron and proton energy, respectively.

    Since we focus on the contribution which has the coefficient $(\alpha/v)^n$, 
    where $v=|\bm{q}|/q^0$ is the relative velocity between proton and electron in this frame, 
    we set $0\le|\bm{q}|\ll m_e \ll m_p$.
    The diagram is `dangerous \cite {Weinberg}' only for $|\bm{k}_i|\to0$.  
    The term which numerator contains $\bm{k}_i$ is ignored since these terms lower the spherical degree of divergence,  
    and it converges in the limit $|\bm{q}|\to0$.
    eq. (\ref{E1}) becomes
\begin{align} \begin{split}\label{E2}
iM_n \propto&
\int d^3\bm{k}_1
\frac{ q^0}{|\bm{k}_1| }
\frac{-1}{|\bm{k}_1| }
\frac{-2}{ |\bm{k}_1|^2+2 \bm{q}\cdot\bm{k}_1}\\
&\times
\int d^3\bm{k}_2
\frac{ q^0}{|\bm{k}_2| }
\frac{-1}{|\bm{k}_2| }
\frac{-2}{ |\bm{k}_1+\bm{k}_2|^2+2 \bm{q}\cdot(\bm{k}_1+\bm{k}_2)}\\
&\times \cdots.
\end{split} \end{align}

%
    For $|\bm{q}|\not=0$, each integral has a logarithmic divergence at $|\bm{k}_i|=0$.
    This is the ordinary infrared divergence.

    On the other hand, if $|\bm{q}|=0$, the amplitude becomes
\begin{align} \begin{split}
iM_n \propto&
\int d^3\bm{k}_1
\frac{ q^0}{|\bm{k}_1| }
\frac{1}{|\bm{k}_1| }
\frac{2}{ |\bm{k}_1|^2}
\times
\int d^3\bm{k}_2
\frac{ q^0}{|\bm{k}_2| }
\frac{1}{|\bm{k}_2| }
\frac{2}{ |\bm{k}_1+\bm{k}_2|^2}
\times \cdots.
\end{split} \end{align}
    Each $\bm{k}_i$ integral has a linear divergence. 
    This is different from the infrared divergence, and it is caused by the condition $|\bm{q}|=0$.
    From the dimensional analysis, we see that each integral in eq. (\ref{E2}) actually has the factor ${q^0}/{|\bm{q}|}=1/v$.

    One denominator $( |\sum \bm{k}_i|^2+\bm{q}\cdot \sum \bm{k}_i)^{-1}$ leads to one $|\bm{q}|^{-1}$ factor. This energy denominator corresponds to the two-fermion intermediate sate.
    At most, the order in $1/v$ is the same as the order in $\alpha$.

\begin{figure}[t]
  \begin{center}
    \includegraphics[keepaspectratio=true,height=16mm]{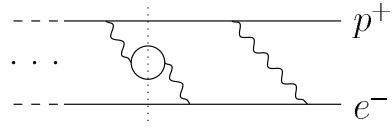}
  \end{center}
  \caption{Fermion loop diagram. At the time indicated by the dotted line, there are four fermion lines. 
  }
  \label{FLoop}
\end{figure}

\begin{figure}[t]
  \begin{center}
    \includegraphics[keepaspectratio=true,height=17mm]{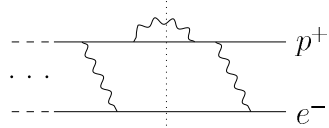}
  \end{center}
  \caption{Fermion self energy. 
  On the dotted line, there are two fermions and one photon, which lower the order in $1/v$.}
  \label{selfE}
\end{figure}

\begin{figure}[t]
  \begin{center}
    \includegraphics[keepaspectratio=true,height=18mm]{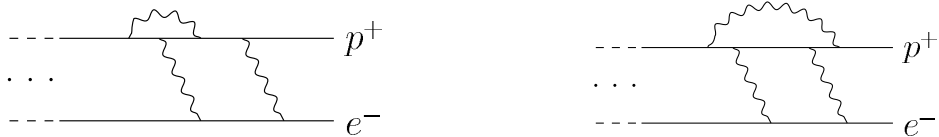}
  \end{center}
  \caption{Vertex corrections. These types of diagrams also include the state which is consisted not only by two fermions but also by some photons.}
  \label{seagul}
\end{figure}

\begin{figure}[t]
  \begin{center}
    \includegraphics[keepaspectratio=true,height=16mm]{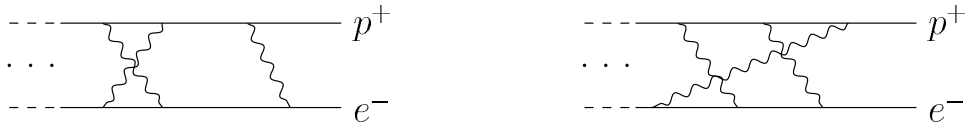}
  \end{center}
  \caption{Crossed ladder diagrams. The situation is the same as the diagrams in figure \ref{seagul}.}
  \label{CL}
\end{figure}

    Next, we consider the diagrams which contain the fermion loop, fermion self energy, vertex corrections, and the crossed ladder parts depicted for instance in figures \ref{FLoop}-\ref{CL}.  
    These diagrams contain less two fermion intermediate states compared to the ladder diagram for the same order in $\alpha$.
    Then, compared to the ladder diagrams, some of the energy denominators of these diagrams depicted for instance in figures \ref{FLoop}-\ref{CL} become
\begin{align} \begin{split}\label{ED}
&\left[m_p+\omega_e(\bm{q})-\{\omega_p(\Sigma\bm{k}_i)+\omega_e(\bm{q}+\Sigma\bm{k}_i)
+\cdots \}\right]^{-1}\\
\simeq& -\left[\frac{2\bm{q}\cdot\sum \bm{k}_i+|\sum \bm{k}_i|^2}{2m_e}
+\cdots \right]^{-1}
\end{split} \end{align}
    instead of the two fermion intermediate sates.
    If ``$+\cdots$" contains the massive particle, this factor does not have a divergence
    since only when the denominator vanish at the endpoint of the range of integration, it may have a divergence.
    For example, this factor is marked by the vertical dotted line in figure \ref{FLoop}.
    Even if $+\cdots$ contains only the photons, $\omega_\gamma(\bm{k}_j)$ (see the vertical dotted line in figure \ref{selfE}), at most eq. (\ref{ED}) is proportional to $1/|\bm{k}_j|$ for $|\bm{k}_i|\to 0$.
    This corresponds to the ordinary infrared divergence.
    This factor does not lead $1/v$ factor.
    Thus, the order in $\alpha$ is larger than the order in $1/v$.
    We ignore these diagrams. 
    Similarly, the diagrams which contain the electromagnetic interactions between the initial and final-state do not contribute to the Fermi function, since this kind of electromagnetic interaction reduce the number of two-fermion intermediate sates.
    This is important for the beta decay of the charged particles.

\begin{figure}[t]
  \begin{center}
    \includegraphics[keepaspectratio=true,height=20mm]{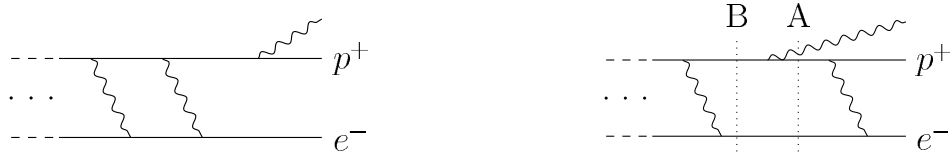}
  \end{center}
  \caption{Bremsstrahlung diagrams. On the dotted line A, there are two fermions and one photon. On the other hand, on the dotted line B, there are only two fermions. }
  \label{brems1}
\end{figure}

    Last, we consider the bremsstrahlung diagrams depicted for example in figure \ref{brems1}.
    Some of the energy denominators become
\begin{align} \begin{split}\label{K}
&\left[m_p+\omega_e(\bm{q})+\omega_\gamma(\bm{k}_j)-\{\omega_p(\Sigma\bm{k}_i)+\omega_e(\bm{q}+\Sigma\bm{k}_i)+\omega_\gamma(\bm{k}_j) \}\right]^{-1}\\
\simeq&
 -\left[\frac{2\bm{q}\cdot\sum \bm{k}_i+|\sum \bm{k}_i|^2}{2m_e} \right]^{-1}.
\end{split} \end{align}
    This factor corresponds to the time marked by the vertical dotted line A in figure \ref{brems1}.
    These denominators lead $|\bm{q}|^{-1}$ factor.
    Some other denominators become
\begin{align} \begin{split}\label{K2}
&\left[m_p+\omega_e(\bm{q})+\omega_\gamma(\bm{k}_j)-\{\omega_p(\Sigma\bm{k}_i)+\omega_e(\bm{q}+\Sigma\bm{k}_i) \}\right]^{-1}\\
\simeq&
 -\left[\frac{2\bm{q}\cdot\sum \bm{k}_i+|\sum \bm{k}_i|^2}{2m_e}-|\bm{k}_j| \right]^{-1}.
\end{split} \end{align}
    This factor corresponds to the time marked by the vertical dotted line B in figure \ref{brems1}.
    These denominators do not lead $|\bm{q}|^{-1}$ factor for $|\bm{k}_i|\to0$.
    Then, the order in $\alpha$ is larger than the order in $1/v$ for the bremsstrahlung diagrams.     
    We ignore these diagrams. 
    After all, only the ladder diagrams contribute to the electromagnetic correction for $n$th order in $\alpha/v$.

\end{document}